	\documentclass[10pt,twocolumn,twoside] {IEEEtran}
	\IEEEoverridecommandlockouts
	\usepackage{amsmath, amssymb, amsthm} 
	\usepackage{graphicx,color}           
	\usepackage{algorithmic}
	\usepackage{hyperref}
	\usepackage{psfrag}
	\usepackage{epstopdf}
	\usepackage{graphicx}
	\usepackage{caption}
	\usepackage{subcaption}
	\usepackage{float}
         \usepackage{multicol}
         \usepackage{pgfplots}
         \usepackage{psfrag} 
         \usepackage{setspace}
\begin{document}
	%
	\title{60 GHz Wireless Link Within Metal 
	Enclosures: Channel Measurements and System Analysis}
	
	\author{Seyran Khademi,~\IEEEmembership{Student Member,~IEEE,}
	Sundeep Prabhakar Chepuri,~\IEEEmembership{Student Member,~IEEE,} 
	Zoubir Irahhauten,~\IEEEmembership{Member,~IEEE,} Gerard 
	J. M. Janssen, \IEEEmembership{Member,~IEEE,}\\ ~and Alle-Jan van der Veen,~\IEEEmembership{Fellow,~IEEE}
	\thanks{This work was supported in part by STW under the FASTCOM project (10551).}
	\thanks{S. Khademi, S. P. Chepuri, G. J. M. Janssen, and A.-J. van der Veen are with the 
	Faculty of Electrical Engineering, Mathematics and 
	Computer Science, Delft University of Technology, The Netherlands.
	Email:~\{s.khademi, s.p.chepuri,g.j.m.janssen,a.j.vanderveen\}@tudelft.nl.}
	\thanks{Z. Irahhauten is with the Mobile Innovation Radio group, KPN, The Netherlands. 
	Email:~zoubir.irahhauten@kpn.com}}
	
	
	\maketitle
	
	\begin{abstract}
	Wireless channel measurement results for 60 GHz within a closed metal cabinet are 
	provided. A metal cabinet is chosen to emulate the environment within a mechatronic system, 
	which have metal enclosures in general. A frequency domain sounding technique is used to measure 
	the wireless channel for different volumes of the metal enclosure, considering both line-of-sight 
	(LOS) and non-line-of-sight (NLOS) scenarios. Large-scale and small-scale characteristics 
	of the wireless channel are extracted in order to build a comprehensive channel model. In contrast to 
	conventional indoor channels at 60 GHz, the channel in the metal enclosure is highly reflective 
	resulting in a rich scattering environment with a significantly large root-mean-square (RMS) delay spread. 
	Based on the obtained measurement results, the bit error rate (BER) performance is evaluated 
	for a wideband orthogonal frequency division multiplexing (OFDM) system.
	\end{abstract}
	
	\begin{IEEEkeywords}
     channel characterization and modeling, frequency domain sounding, 60~GHz measurements, orthogonal 
     frequency division multiplexing (OFDM), root-mean-square (RMS) delay spread.
       \end{IEEEkeywords}
%
	\section{Introduction}\label{sec:I}
	\IEEEPARstart{A}{}dvances in wireless communications have resulted in an ever-increasing desire and 
	demand for high-rate wireless data transfer not only for unlimited access to information and entertainment, 
	but also for industrial usage and sensor network applications. Current wireless technology does not 
	meet the data rates offered by wired standards like gigabit Ethernet. Recently, the unlicensed multi-GHz 
	spectrum available around 60 GHz has gained a lot of interest, as it has the ability to support short-range 
	high data rates in the order of Gbps~\cite{Daniels,Heath,IEEE80215c}. 
	
	In this paper, we consider high-rate short-range wireless
	communication for mechatronic systems.  The demand for high data
	rates (peak data rate up to a few tens of Gbps) comes from a
	substantial number of high rate sensors and actuators, but also
	from a very tight control loop in which little time is allocated
	for the actual communications.  Application areas include
	communication within automobiles, satellites, aircrafts or
	industrial machineries.  These environments with metal
	enclosures are highly reflective, and the resulting ``long''
	wireless channels make wireless communications very challenging.

	For outdoor and indoor 60~GHz wireless systems, an
	extensive literature exists related to channel measurements and characterization.
	For outdoor scenarios, the conventional 60 GHz wireless channel exhibits increased free space path-loss with up to 
	$14 \, \mathrm{dB/km}$ atmospheric absorption ~\cite{Daniels,Heath}.
	This reduces the multipath effect but also makes non-line-of-sight (NLOS) communication more 
	difficult \cite{Smulders}. 
	For indoor scenarios, \cite{Hao} studied the characteristics of 60~GHz channels 
	both in space and time for short-range broad-band wireless system design.
	The channel multipath structure is provided based on angle of arrival (AOA) and time of arrival (TOA)
	components which is helpful for incorporation in spatial signal processing algorithms.
	Further, \cite{Moraitis} focuses on the small-scale nature of the channel in different indoor environments. 
	The task group TG3c \footnote{The IEEE P802.15.3 
	    working group for wireless personal area networks (WPANs) introduced a task-group, 
	    referred to as TG3c.} 
	produced a channel modeling document \cite{IEEE80215c}
	reporting large-scale and small-scale channel parameters for a variety of indoor environments
	related to WPAN systems.
        Another comprehensive survey on 60~GHz indoor channel models is presented in \cite{Smulders}.
	However, to the best of our knowledge, there is no report on channel measurements and modeling within a metal 
	enclosure, as studied in the present paper.
	\subsection{Why is the 60~GHz band interesting for mechatronic systems?}
	Due to spectrum scarcity, the 60 GHz band has recently gained a lot of attention. Unlike the crowded 
	2.4~GHz and other low frequency ``free'' bands,  unlicensed and unoccupied spectrum is abundantly available 
	at 60~GHz. 
	This motivation does not directly apply to mechatronic systems within closed metal enclosures, in which
	transmissions do not interfere with other existing communication 
	systems. However, several properties do make communication at 60~GHz within metal enclosures 
	interesting, as follows:
	\begin{itemize}
	\item  \textbf{Small antenna size}: Higher frequencies lead to smaller size antennas, which enables us to put more antennas 
	in a small area, and perhaps even integrate all of them on a chip \cite{AntennaOnChip}. The small size 
	of the antennas makes multiple antennas more feasible for short-range wireless communications. 
	Recently, massive-MIMO systems have been proposed in~\cite{Larsson} which could be a milestone in 
	boosting the data rate in wireless systems. Such systems benefit from rich scattering environments. Thus,
	metal enclosures combined with 60~GHz technology and very-large MIMO techniques should enable 
	high data rates comparable to wired systems~\cite{Seyran}.
	\vspace{0.1cm}
	\item \textbf{Physical available bandwidth}: More physical
	bandwidth is available when we go to higher carrier frequencies.
	E.g., the maximum physical bandwidth that can be used at $f_c=
	2.4$~GHz is limited to $2 f_c = 4.8$~GHz, and the usable
	bandwidth is often less than this.  E.g., in highly dispersive
	channels (like metal enclosures), the usable bandwidth is
	reduced due to the large number of nulls in the 
	channel frequency response.
	 \end{itemize}
	\subsection{Contributions and outline}
	For signal processing and physical layer system design, a fair parameterized model of the propagation environment 
	is essential. This includes the delay spread in time, the Doppler
	spread in frequency, a shadowing and path-loss model, etc. In this paper, a 60~GHz wideband wireless channel 
	model based on frequency domain sounding within a metal cabinet is proposed.
	This was considered to emulate the environment of a mechatronic system, the main motivation for this work. 
	We present channel measurement results and their analysis for different sizes of the metal enclosure, for both 
	line-of-sight (LOS) and non-line-of-sight (NLOS) scenarios. A comprehensive channel model is 
	developed to attribute both the small-scale and large-scale channel characteristics. 
	
	To understand the challenges related to high data rate 
	wireless communications in such environments, we evaluate the bit-error-rate (BER) performance of a 
	wideband orthogonal frequency division multiplexing (OFDM) system for the 
	measured channels, and compare these to a conventional Rayleigh fading channel. We also provide 
	a system design based on link budget calculations for an example high data-rate system.
	
	The remainder of this paper is organized as follows. In Section \ref{sec:II}, we discuss the measurement set-up and 
	explain the measurement procedure. In Section \ref{sec:III},  we provide details regarding data processing to extract 
	parameters required for channel modeling. Based on these parameters, large-scale and small-scale channel 
	models are presented in Sections \ref{sec:IV} and \ref{sec:V}, respectively. In Section \ref{sec:VI}, we compare the measured 
	channel with the Saleh-Valenzuela (SV) channel model suggested for the IEEE 802.15 standard. 
	The system analysis for a wireless link is discussed in Section \ref{sec:VII}. Final remarks are made in section \ref{sec:VIII}.

	\section{Measurement set-up and procedure} \label{sec:II}
	In this section, the channel measurement procedure is explained.  We also 
	provide details on the equipment used for the measurements. 
	\subsection{Frequency domain sounding technique}
\begin{figure}
	\centering
	\psfrag{B}{$B_w$}
	\psfrag{G}{frequency}
	\psfrag{A}{$\Delta f_s$}
	\psfrag{C}{$f_{min}$}
	\psfrag{D}{$f_{max}$}
	\psfrag{f}{$f_{c}$}
	\includegraphics[height=1.5in, width=3.4 in]{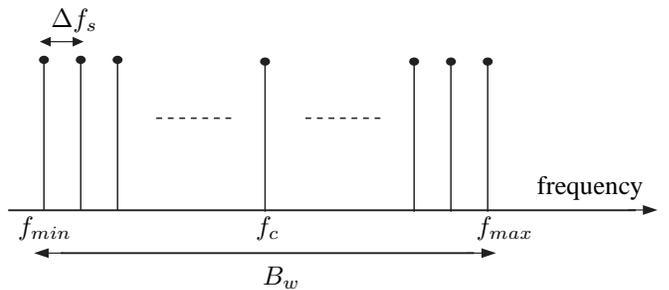}
	\caption[Measurement Setting 2]{Frequency domain sounding technique.}
	 \label{fig:FD}
	\end{figure}

	Channel characterization can be performed in either time domain or frequency domain \cite{Parsons}.
	In the measurements 
	provided in this paper, a frequency domain sounding technique is used. The scattering parameters 
	(i.e. $S_{11}$, $S_{12}$, $S_{21}$, $S_{22}$) are measured using a vector network analyzer (VNA) by 
	transmitting sinusoidal waves at discrete frequencies, 
	wherein the full bandwidth of interest is sampled using an equal frequency spacing, as depicted in Fig.~\ref{fig:FD}. 
	For each of these sinusoids, the magnitude and phase of the transmitted and received signals are measured.
	The frequency spacing $\Delta f_s$ and the scanned bandwidth $B_w$ determine the resolution $\tau_{res}$ of 
	the captured multipaths and the maximum measurable excess delay $\tau_{max}$, as
	\begin{equation}
	 \Delta f_s=\frac{B_w}{N_s},\quad \tau_{max}=\frac{1}{\Delta f_s},\quad \tau_{res}=\frac{1}{ B_w} \,,
	\end{equation}
	where $N_s$ is the number of transmitted sinusoidal waves.
	
	The frequency domain $S_{21}$ parameter is generally referred to as the channel frequency response (CFR).
	The channel impulse response (CIR) is obtained from the measured CFR by taking the inverse fast Fourier transform (IFFT).
	A Hann window is applied to reduce the effect of side lobes.
	\subsection{Measurement set-up}
	The measurement bandwidth is set to $B_w=5$~GHz, and the channel is sampled
	from $57$~GHz to $62$~GHz at $N_s=12001$ frequency points.  This
	results in a frequency spacing of $\Delta f_{s}=0.416$~MHz, so that the
	time resolution is $\tau_{res} =\frac{1}{B_w}=0.2$~ns and
	the maximum measurable excess delay is
	$\tau_{max}=2400\,$~ns.
	The channel frequency response is measured using a PNA-E series microwave
	vector network analyzer (VNA) E8361A from Agilent.  
	An intermediate frequency bandwidth of $B_{IF}=50$~Hz is chosen to
	reduce the noise power within the measurement band, which
	improves the dynamic range. This is the receiver bandwidth for single sinusoid in the VNA; 
	the smaller is IF bandwidth the larger the signal to noise ratio.
	Also each measurement is repeated 50 times to further average out the noise.  

\begin{figure}
	\centering
	\psfrag{A}{\small metal cabinet}
	\psfrag{B}{\small Tx antenna}
	\psfrag{C}{\small Rx antenna}
	\psfrag{D}{\small Agilent PNA$-E8361A$}
	\psfrag{E}{\small $s_{out}$}
	\psfrag{F}{\small $s_{in}$}
	\psfrag{G}{\small GPIB}
	\psfrag{H}{\small PC}
	\psfrag{I}{\small 60 GHz}
	\psfrag{J}{\small \small PA}
	\psfrag{K}{\small DC}
	\psfrag{L}{\small supply}
	\includegraphics[height=2.1in,width=3.4 in]{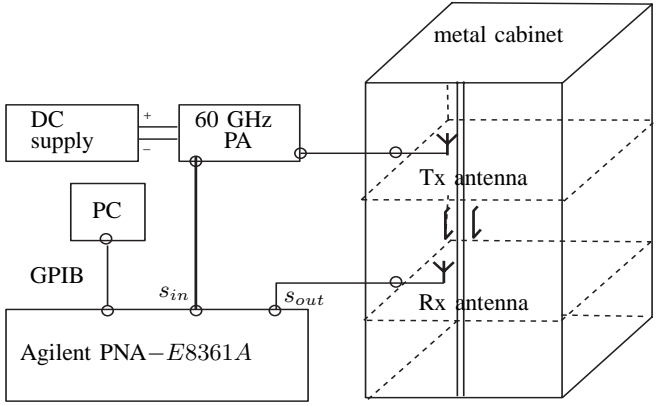}
	\caption[Measurement Setting 1]{Measurement setup for channel sounding inside the metal cabinet.}
	 \label{fig:PNA}
	\end{figure}

	Due to the losses inside the VNA and  $60$~GHz co-axial cables, the measured signal at the 
	receiver is weak. A $60$~GHz solid state power amplifier (PA) from QuinStar Inc.\ (QGW-50662030-P1) 
	was used to compensate for the losses and to improve the dynamic range. An illustration of the measurement 
	set-up is provided in Fig.~\ref{fig:PNA}.
	\begin{figure}
	\centering
	\psfrag{H-plane}{\small H-plane}
	\psfrag{E-plane}{\small E-plane}
	\includegraphics[height=2.3in,width=2.4 in]{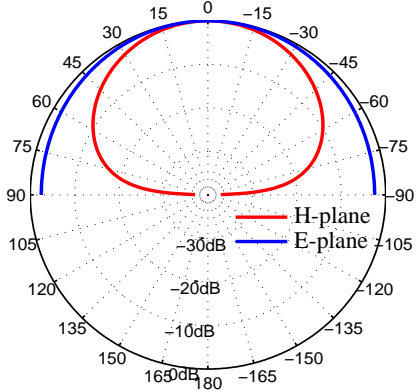} 
	\caption[Radiation Pattern for Open Waveguide]{Field radiated by the $\text{TE}_{10}$ mode in open 
	waveguide antenna with respect to $\theta$ angle.}
	\label{fig:Rad1}
	\end{figure}
	As transmit and receive antennas, we used
	two identical open waveguide antennas for the 50-75~GHz frequency band, with aperture size 
	$3.759 \times 1.880$~mm$^2$.
	The beam pattern of the antennas 
	is shown in Fig.~\ref{fig:Rad1}. The gain of the open waveguide antenna is about $4.6$~dBi 
	as calculated using \cite{antenna1} 
	\begin{equation}\label{eq:antenna gain}
	G=\frac{4\pi A_e (f_c)^2}{c^2},
	\end{equation}
	where $A_e$ is the effective aperture and $c$ is the speed of light.
	\subsection{Measurement procedure}
	To investigate the channel behavior within the metal cabinet, we considered three scenarios.

	{\it Scenario~1} is a LOS scenario where we used a metal enclosure of dimension $100 \times 45 \times 45~\mathrm{cm}^3$. 
	{\it Scenario~2} is a LOS scenario with 
	a metal enclosure of a larger dimension, i.e., $100 \times 45 \times 180~\mathrm{cm}^3$.
	{\it Scenario~3} is an NLOS scenario with the same larger enclosure.
	Note that the volume of the metal enclosure in {\it scenario~2} and {\it scenario~3} is four 
	times larger than the volume of the metal enclosure used for {\it scenario~1}. To block the LOS path, and create 
	the NLOS scenario, a $50 \times 45$~cm metal separation plate is used in {\it scenario~3}. 
	
	The transmit and receive antennas were placed on a styrofoam (polystyrene) sheet, which acts as vacuum for 
	radio waves and has a negligible effect on the channel behavior. The transmit and receive antennas were supported 
	using metal clamps. The co-axial cables were drawn into the metal cabinet by means of small holes which are 
	just sufficiently large to pass the cable. 
	
	For all scenarios, the location of the transmit antenna was kept fixed. The channel was measured at various locations in 
	$3$ dimensions, i.e., $x,y,z$ axes, as specified
	in Table \ref{Table1}. This produced 96, 96 and 72 receiver locations for scenario 1, 2 and 3,
	respectively.
	\begin{table}
	\centering
	    \begin{tabular}{| l | l | l | l |}
	    \hline
	     & $x$-axis & $y$-axis & $z$-axis \\ \hline
	    Scenario 1& 15-85 cm; 8 steps & 5-30 cm; 6 steps& 15, 30 cm \\ \hline
	    Scenario 2 & 15-85 cm; 8 steps& 5-30 cm; 6 steps& 35, 140 cm \\ \hline
	    Scenario 3& 15-40 cm; 6 steps& 5-30 cm; 6 steps& 35, 140 cm \\
	    \hline
	    \end{tabular}
	    \caption{Receive antenna co-ordinates.} \label{Table1}
	\end{table}
	
	In {\it scenario~1} and {\it scenario~2} the transmit antenna was fixed at co-ordinate $(x_t,y_t,z_t)=(65,15,0)$ cm, and 
	in {\it scenario~3}, for the NLOS case the transmit antenna was located at 
	 $(x_t,y_t,z_t)=(15,15,0)$ cm. 

	%
	\section{Data processing}\label{sec:III}
	\subsection{Inverse filtering technique}
	Post-processing of the data is required to extract the channel impulse response (CIR) from the measured 
	frequency domain signals. Prior to the IFFT, we cancel the antenna and instrument responses using an
	inverse filtering technique \cite{Siamarou,Ghassemzadeh}, briefly explained as follows.
	
	Let $x(t)$ be the transmitted signal, which is impaired by the measurement system and the antennas. 
	The received signal $r(t)$ is modeled as
	\begin{equation}
	\label{eq:InverseFiltering0}
	 r(t)=x(t)*h_{tx}(t)* h_{sys}(t)* h(t) * h_{rx}(t),
	\end{equation}
	where $h_{tx}(t)$ and $h_{rx}(t)$ are the impulse responses of the transmit and receive 
	antennas, $h_{sys}(t)$ is the transfer function of the measurement system and $h(t)$ is the CIR of interest.
	
	The CIR for free space without reflections or obstructions
	consists of a single LOS path, parametrized by an attenuation
	and a simple delay equal to the time-of-flight of the signal
	between the transmit and receive antenna.  We can make a
	recording of the received signal at a known reference distance
	in free space, and after time gating obtain a reference signal $r_{fl}(t)$, modeled as
	\begin{equation}\label{eq:InverseFiltering1}
	r_{fl}(t)\approx x(t) * h_{tx}(t) * h_{sys}(t) * h_{rx}(t)\,,
	\end{equation}
	so that
	\begin{equation}\label{eq:InverseFiltering2}
	r(t)\approx r_{fl}(t) * h(t).
	\end{equation}
	More specifically, $r_{fl}(t)$ in \eqref{eq:InverseFiltering1}
	absorbs the effect of the antennas and the system (this is not
	entirely accurate as the directionality of the antennas is
	ignored).  The CIR is obtained from \eqref{eq:InverseFiltering2} via inverse filtering.
	Equivalently, in frequency domain, we can obtain the CFR $H(f)$ by
	\begin{equation}\label{eq:Deconv0}
	H(f)=\frac{R(f)}{R_{fl}(f)} \,.
	\end{equation}
	The CIR is then obtained by taking the (windowed) IFFT of $H(f)$ and
	correction for the delay and attenuation (normalization). 
	
	We have obtained a reference LOS signal $r_{fl}(t)$ by placing the
	transmitter and receiver at a distance of $25$~cm outside the
	metal cabinet (free space). The LOS path was retrieved by
	windowing the measured signal and truncating it after 50 ns (time gating), 
	so as to remove noise and multipaths beyond the direct line of sight.

	\begin{figure}
	    \centering
	    \psfrag{P}{\small reference CIR ($d_0=25$~cm)}
	    \psfrag{L}{\small sample CIR in cupboard ({\it scenario1})}
	    \psfrag{C}{\small sample CIR after inverse filtering}
	    \psfrag{B}{\small power (dB)}
	    \psfrag{A}{\small time (ns)}
	    \includegraphics[width=3.45 in,height=2in]{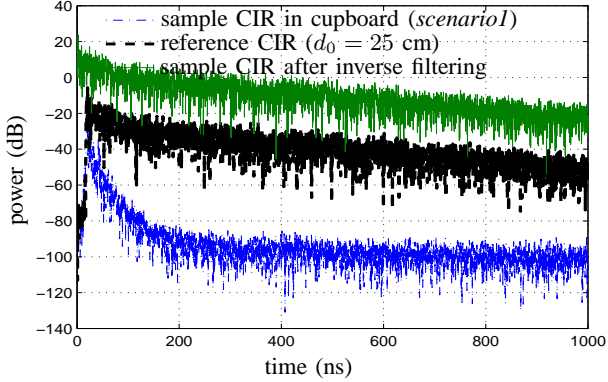}
	    \caption{Sample CIR from {\it scenario~1}  before and after inverse filtering 
	    with Tx-Rx $d=33$~cm apart, compared to the measured reference CIR (free space,
	    Tx-Rx $d_0=25$~cm apart). }
	     \label{fig:CIR}
	 \end{figure}
	\begin{figure}
	    \centering
	    \psfrag{Y}{\small power (dB)}
	    \psfrag{X}{\small frequency (GHz)}
	    \psfrag{sample CFR after inverse filtering}{\small sample CFR after inverse filtering}
	    \psfrag{sample CFR in cubpoard (scenario1)}{\small sample CFR in cupboard ({\it scenario1})}
	    \psfrag{reference CFR (25 cm)}{\small reference LOS CFR ($d_0=25$~cm)}
	    \includegraphics[width=3.45 in,height=2in]{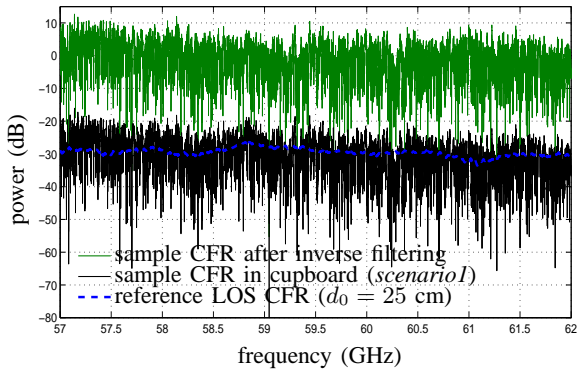}
	    \caption{Sample CFR from {\it scenario~1} before and after inverse filtering and reference CFR (free 
	space).}
	     \label{fig:CFR}
	 \end{figure}
	\begin{figure}
	\centering
	\psfrag{threshold (dB)}{\small threshold (dB)}
	\psfrag{normalized power}{\small normalized power}
	\includegraphics[width=3.4 in,height=2in]{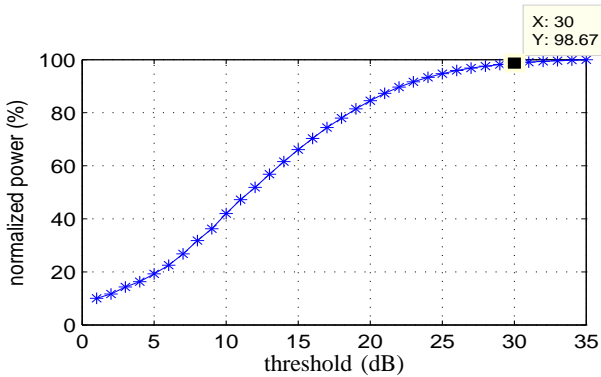}
	\caption{Average received power for different thresholds in {\it scenario~1}.}
	 \label{fig:ThreshVsPower}
	\end{figure}

	In Fig.~\ref{fig:CIR}, we show a sample measurement CIR within
	the metal cabinet ({\it scenario~1}, at 33 cm, with the transmitter
	and receiver at different heights), as well as the measured free
	space signal.  
	The figure illustrates the effect of the metal enclosure on the received
	multipaths which are above the noise floor even up to 1000~ns.

	Fig.~\ref{fig:CFR} shows the original frequency domain response
	(CFR) of the sample measurement from {\it scenario~1}, and the
	frequency domain signal $R_{fl}(f)$ of the truncated reference
	measurement $r_{fl}(t)$.  The effect of inverse filtering can be
	observed in the calibrated plot where the sample CFR is
	normalized by $R_{fl}(f)$.  

	\subsection{Normalization and thresholding}\label{sec:threshold}
	For model parameters that do not depend on the absolute power (i.e.,
	the small-scale channel model considered in Sec.\ref{sec:V}), we
	have normalized the received signal to have a maximum value at
	$0$ dB.  The dynamic range of the received signal is in the
	order of $70$~dB, considering the noise level at $-70$~dB after
	normalization.

	For estimating statistics for individual path parameters, it is
	useful to truncate the duration of the channel.  We apply a
	threshold level taking into account the noise level, the amount
	of total received power and the relevant multipath components
	\cite{Zoubir2,Zoubir3}.

\begin{figure}
	\centering
	\psfrag{time (ns)}{\small time (ns)}
	\psfrag{normalized CIR (dB)}{\small normalized CIR (dB)}
        \psfrag{threshold}{\small threshold}
        \psfrag{selected paths}{\small selected paths}
        \psfrag{sample CIR}{\small sample CIR}
	\includegraphics[width=3.4 in,height=2 in]{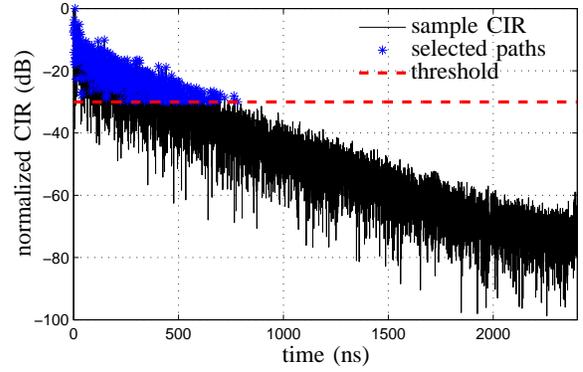}
	\caption[Threshold Setting]{Sample CIR with 30~dB threshold and received paths for {\it scenario 1}}
	 \label{fig:ScTr}
	\end{figure}

	From Fig.~\ref{fig:ThreshVsPower} it can be observed that by setting a threshold at 30~dB below the 
	strongest path, more than $98\%$ of the total power is captured in {\it scenario~1}. This threshold is still well above the noise level. 
	As an illustration, Fig.~\ref{fig:ScTr} shows a normalized received CIR with a threshold at $-30$~dB.
	The duration of this channel is still about 800 ns.
	%

	%
	\section{Large-scale channel model: Path-loss}\label{sec:IV}
	The large-scale channel model, specifically the path-loss model, is essential for any wireless system design to 
	calculate its link budget. For a conventional channel (outdoor or indoor), the path-loss model suggests that the 
	average received power decreases exponentially with increasing distance between transmitter and 
	receiver. This is generally expressed in logarithmic scale as
	\begin{equation}\label{eq:Pathloss}
	 P_L(d)_{dB}=P_L(d_0)_{dB}+\alpha10\log_{10}(\frac{d}{d_0})+X_{\sigma}.
	\end{equation}
	where
	$P_L(d)_{dB}$ is the average received power at a distance $d$~(m) relative to a reference 
	distance $d_0$~(m), $\alpha$ represents the path-loss
	exponent, and $X_{\sigma}$ is a zero-mean Gaussian random variable with standard deviation $\sigma$ 
	reflecting the attenuation (in dB) caused by flat fading (shadowing or slow fading).
	\begin{figure}
	\begin{subfigure}{0.5\textwidth}
	\centering
	\psfrag{distance d (m)}{\small distance d (m)}
	\psfrag{path loss (dB)}{\small path loss (dB)}
	\psfrag{measured data scenario1}{\small measured data {\it scenario~1}}
	\psfrag{A}{\small $P_L(d)=54.711+0.021\cdot 10\log10(d)$}
	\psfrag{measured data scenario2}{\small measured data {\it scenario~2}}
	\psfrag{B}{\small $P_L(d)=53.439+0.004\cdot 10\log10(d)$}
	\psfrag{measured data scenario3}{\small measured data {\it scenario~3}}
	\psfrag{C}{\small $P_L(d)=54.116+0.002\cdot 10\log10(d)$}
	\includegraphics[width=3.4 in,height=2.7in]{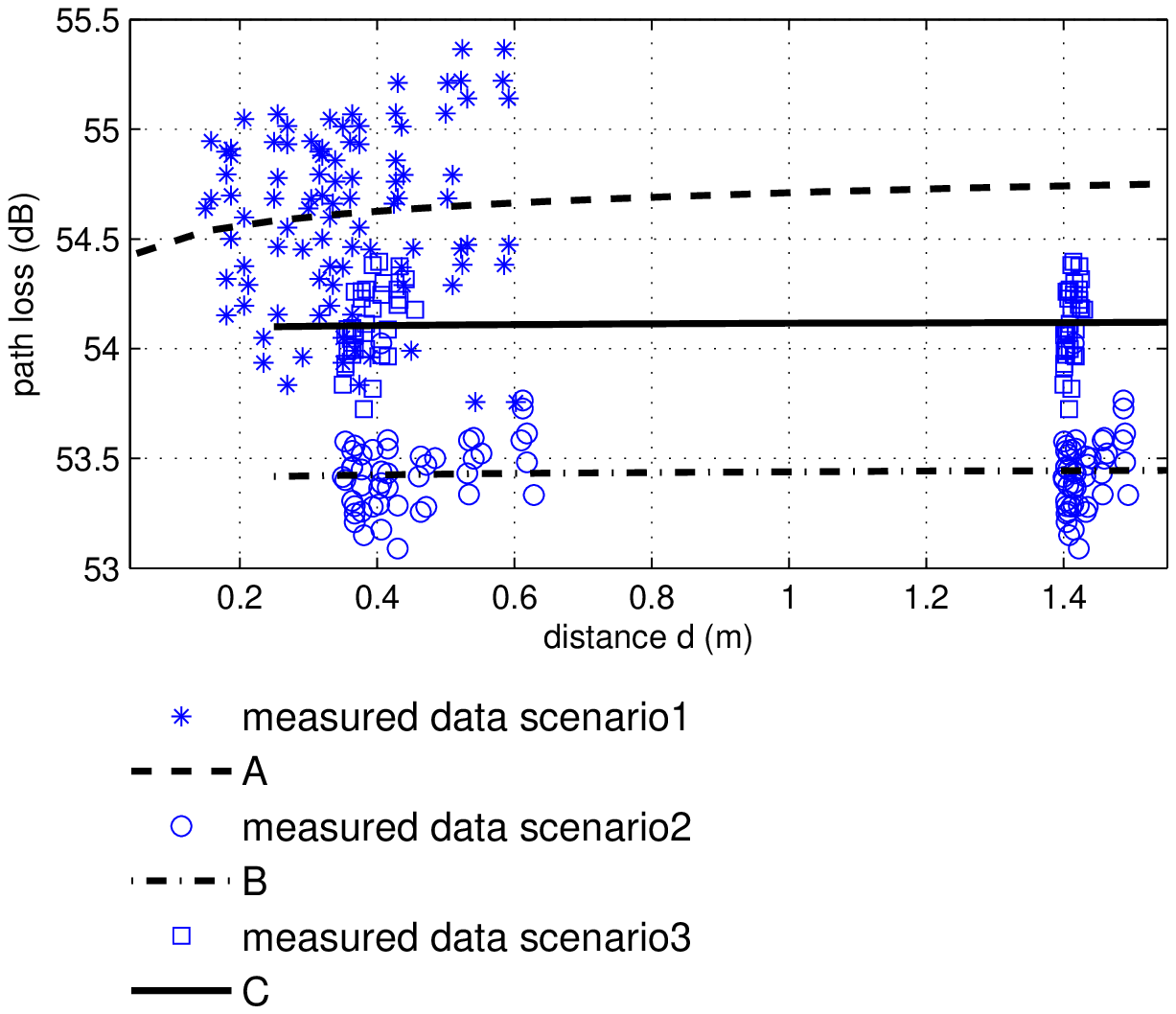}
	\caption{Path-loss as function of distance.}
	 \label{fig:PathLossModel}
	\end{subfigure}
	\begin{subfigure}{0.5\textwidth}
	\centering
	\psfrag{PDF}{\small PDF}
	\psfrag{path loss variation (dB)}{\small path loss variation (dB)}
	\psfrag{measured data scenario1}{\small measured data {\it scenario~1}}
	\psfrag{A}{\small normal fitting: ($\mu,\sigma$)=(0,0.39)}
	\psfrag{measured data scenario2}{\small measured data {\it scenario~2}}
	\psfrag{B}{\small normal fitting: ($\mu,\sigma$)=(0,0.17)}
	\psfrag{measured data scenario3}{\small measured data {\it scenario~3}}
	\psfrag{C}{\small normal fitting: ($\mu,\sigma$)=(0,0.16)}

	\includegraphics[width=3.4 in,height=2.7in]{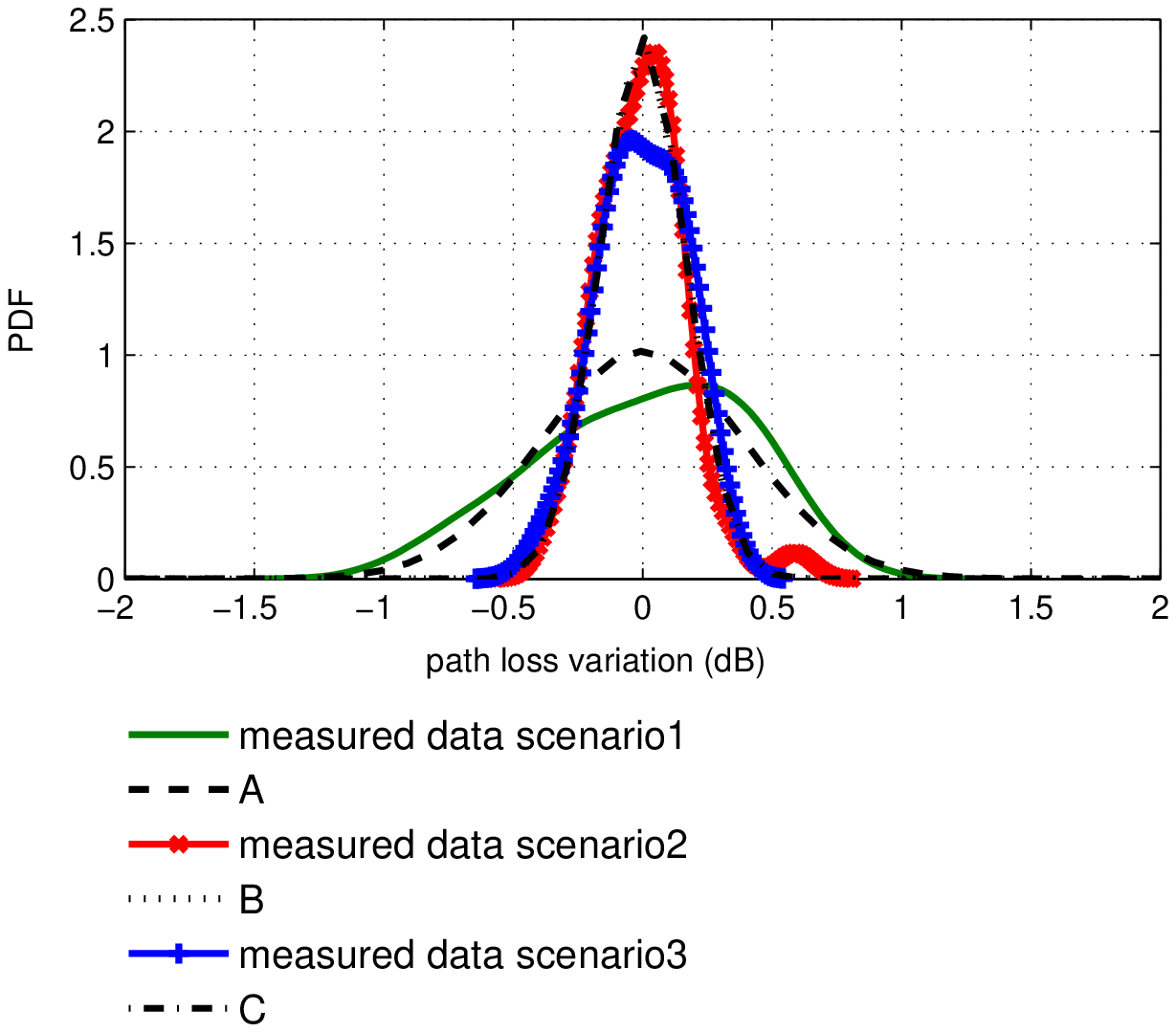}
	\caption{PDF of the path-loss variation $X_\sigma$.}
	 \label{fig:pdfPathlossV}
	\end{subfigure}
	\caption{Path-loss model parameter estimation.}
	\end{figure}

	Using the measurements of the received power for different distances between the transmit and receive 
	antennas, we can estimate the path-loss exponent $\alpha$. 
	Accordingly, for each measurement the distance related path loss term ($P_t-P_r$) is calculated as shown in
	Fig.~\ref{fig:PathLossModel} which shows that the path-loss exponent $\alpha$ is very small (around $0.02$-$0.002$), 
	suggesting that in such a closed metal 
	environment there is nearly no loss in the received power as function of distance.
	In comparison, for narrowband indoor systems some measurements
	have reported a path-loss exponent of $\alpha$ in the range
	$1.6-6$ \cite{Rappaport}, and NLOS wireless personal area
	network (WPAN) measurements resulted in $\alpha$ in the range
	$0.04-0.09$ \cite{Zoubir1,Zoubir2}.  According to the Friis
	formula, the path-loss exponent for conventional indoor
	environments should be larger for transmissions at $60$~GHz
	compared to lower carrier frequencies.

	The ideal metal enclosed environment acts as a
	semi-conservative physical system where the only sources of
	absorptions are the antennas.  The waves keep bouncing back and
	forth, and when the distance between the antennas is incremented
	the received power does not fluctuate because most of the energy
	reaches the receive antenna either directly or as multipath reflection in the metal cabinet.

	Fig.~\ref{fig:pdfPathlossV} shows the probability density
	function (PDF) of $X_\sigma$, i.e., the fluctuation of the path-loss
	around the regression line in Fig.~\ref{fig:PathLossModel}.  It
	is seen that the PDF approximately follows a normal
	distribution, with a standard deviation of $0.16$--$6$ dB.
	Among the considered scenarios, the NLOS case ({\it scenario~3})
	shows the smallest variation, and this is due to the larger
	distances (volume) and the obstructed LOS path.

	%
	\section{Small-scale Channel Model}\label{sec:V}
	The path-loss model describes the channel behavior in
	large-scale in terms of the attenuation as function of distance.
	The channel can be further characterized by its small-scale
	properties caused by reflections in the environment, which are
	modeled as multipath components.  The $n$th multipath component
	is described by a power $a_n^2$ and arrival time $t_n$.
	Multipath leads to small-scale fading (variations over short
	distances due to constructive and destructive additions).  The
	most important model parameters that describe small-scale
	variations are the RMS delay spread, the time decay constant,
	and the PDF of the multipath arrival times.

	\subsection{RMS delay spread (RDS)} %
	\begin{figure}
	\begin{subfigure}{0.5\textwidth} \centering
	\psfrag{Y}{\small number of paths}
	\psfrag{threshold (dB)}{\small threshold (dB)}
	\psfrag{A}{\small {\it scenario~1}}
	\psfrag{B}{\small {\it scenario~2}}
	\psfrag{scenario3 (NLOS)}{\small {\it scenario~3} (NLOS)}

	\includegraphics[width=3.4 in,height=2in]{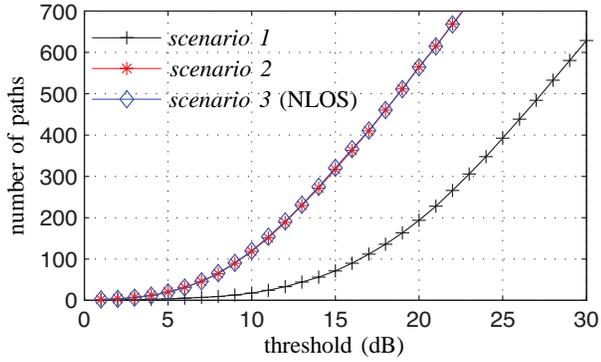}
	\caption{Number of received paths}
	 \label{fig:NrPathSvsTr} \end{subfigure}
	\begin{subfigure}{0.5\textwidth} \centering
	\psfrag{Y}{\small mean RDS (ns)}
	\psfrag{threshold (dB)}{\small threshold (dB)}
	\psfrag{A}{\small {\it scenario~1}}
	\psfrag{B}{\small{\it scenario~2}}
	\psfrag{scenario3 (NLOS)}{\small {\it scenario~3} (NLOS)}
	\includegraphics[width=3.4 in,height=2in]{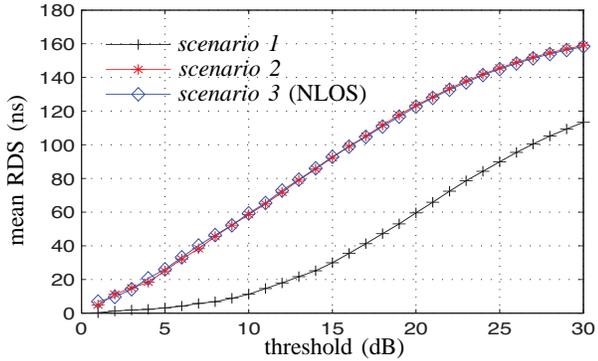}
	\caption{mean RDS}
	 \label{fig:RDSvsTr} \end{subfigure}
	\caption[Number of paths and RDS vs Threshold]{Number of received paths and RDS for different thresholds.}
	\label{fig:RDS} \end{figure} 
	Delay spread describes the time
	dispersion effect of the channel, i.e., the distribution of the
	received power in time.  A large delay spread causes severe
	intersymbol interference (ISI) and can deteriorate the system
	performance.  The RMS delay spread (RDS) is a commonly used
	parameter to characterize this effect \cite{Goldsmith}.  The RDS
	is obtained by first estimating the individual path parameters
	$\{(a_n^2,t_n)\}$ for each observation, and then computing	
    \[
    t_{rms}=\sqrt{ \bar{t^2}- {(\bar{t})}^{2}},\quad
    \bar{t^{\beta}}=\frac{\sum_{n=1}^{N}\,a_n^2 t_n^{\beta}}{\sum_{n=1}^{N}\,a_n^2},
    \]  
        where $\bar{t}$, $\bar{t^2}$ and $\bar{t^{\beta}}$ are the first, second and $\beta$ moment 
        of the power delay profile, respectively. 
        
	Fig.~\ref{fig:NrPathSvsTr} shows the number of received paths
	for different power thresholds.  As expected, the number of
	received paths ($N$) increases with increasing threshold level.  In
	the same way, the RDS increases as the number of collected paths
	increases (Fig.~\ref{fig:RDSvsTr}).  At a threshold of 30 dB, the
	curves saturate and we used the corresponding value as the
	estimated RDS.

	\begin{figure}
	\centering
	\psfrag{CDF}{\small CDF}
	\psfrag{rms delay spread (ns)}{\small rms delay spread (ns)}
	\psfrag{measured RDS scenario1}{\small measured data  {\it scenario1}}
	\psfrag{A}{normal fitting: ($\mu,\sigma$)=(113.4, 12.1)}
	\psfrag{measured RDS scenario2}{\small measured data {\it scenario2}}
	\psfrag{B}{normal fitting: ($\mu, \sigma$)=(159.1, 5.1)}
	\psfrag{measured RDS scenario3}{\small measured data  {\it scenario3}}
	\psfrag{C}{normal fitting: ($\mu,\sigma$)=(158.3, 4.4)}

	\includegraphics[width=3.4 in,height=3 in]{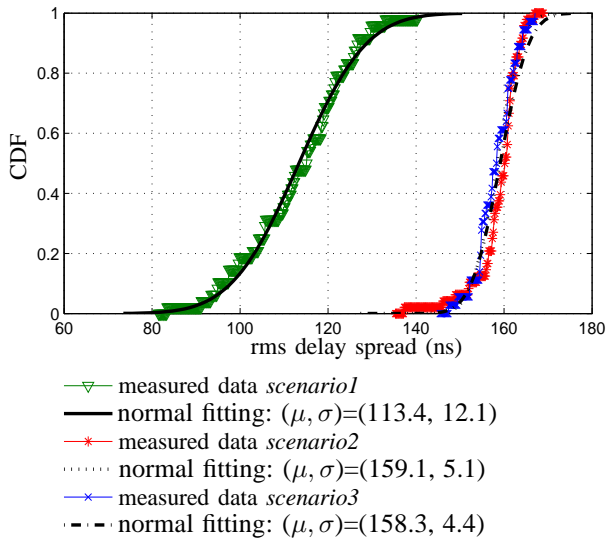}
	\caption[CDF of RDSs]{Cumulative distribution function of RDS.}
	 \label{fig:CFRcomparison}
	\end{figure}
	Fig.~\ref{fig:CFRcomparison} shows the cumulative distribution function (CDF) of the estimated RDS values for each scenario.
	The figure also shows the fit to a normal distribution. The 
	mean values of the normal distribution, obtained after fitting, reveals the average length of the channel. 
	We obtained values of $113.4$ ns ({\it scenario 1}), $159.1$ ns ({\it scenario 2} ) and $158.3$ ns ({\it scenario 3}).
	These mean RDS values are significantly larger than conventional indoor channels, which 
	are typically between $4-21$~ns.
	These large values will impact the signal processing and system
	design within such environments, e.g., the channel equalization
	and residual inter block interference (IBI) after equalization,
	and hence, the achievable data rates.

	Note that the estimated mean RDS is almost the same for {\it
	scenario~2} and {\it scenario~3}, which shows that there is a
	clear relation between the volume of such metal enclosures and
	RDS, independent of LOS and NLOS cases.

	\subsection{Time decay constant}
	Most current IEEE standard channel models are based on the 
	Saleh-Valenzuela (SV) model \cite{SV1,SV2}. In this model, the multipaths are considered as a number of rays 
	arriving within different clusters, and separate power decay constants are defined for the rays and the clusters.
	
\begin{figure}
	\begin{subfigure}{0.4\textwidth}
	\centering
	\psfrag{Y}{$\log ({a_n^2}/{a_0^2}) $}\psfrag{time delay (ns)}{\small time delay (ns)}
	\psfrag{measured data}{\small measured data}\psfrag{fitted model}{\small fitted model: $\gamma_k=170.916$}
	\includegraphics[width=3.2 in,height=1.5in]{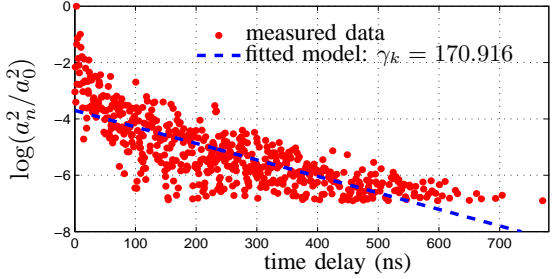}
	\caption{sample measurement in {\it Scenario 1}}
	 \label{fig:Gamma1}
	\end{subfigure}
	\begin{subfigure}{0.4\textwidth}
	\centering
	\psfrag{Y}{$\log ({a_n^2}/{a_0^2}) $}\psfrag{time delay (ns)}{\small time delay (ns)}
	\psfrag{measured data}{\small measured data}\psfrag{fitted model}{\small fitted model: $\gamma_k=204.094$}
	\includegraphics[width=3.2 in,height=1.5in]{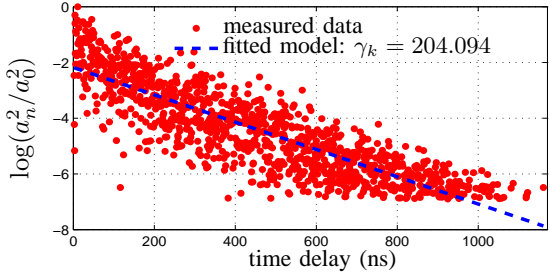}
	\caption{sample measurement in {\it Scenario 2}}
	 \label{fig:Gamma2}
	\end{subfigure}
	\begin{subfigure}{0.4\textwidth}
	\centering
	\psfrag{Y}{$\log ({a_n^2}/{a_0^2}) $}\psfrag{time delay (ns)}{\small time delay (ns)}
	\psfrag{measured data}{\small measured data}\psfrag{fitted model}{\small fitted model: $\gamma_k=189.632$}
	\includegraphics[width=3.2 in,height=1.5in]{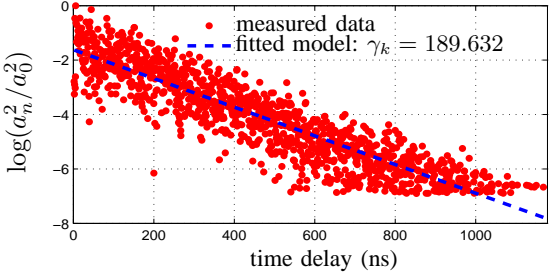}
	\caption{sample measurement in {\it Scenario 3}}
	 \label{fig:Gamma3}
	 \end{subfigure}
	  \caption{LS fit for time decay constant is leading to an estimated  $\gamma_k$ for each measurement.}
	  \label{fig:Gamma}
	\end{figure}

	Our measurement results do not show that the multipath components 
	form clusters. A physical justification comes from the fact that multipath reflections are coming from 
	the (same) walls.
	In this case, the average power delay profile (PDP) is defined by only one 
	decay parameter $\gamma$ rather than the common SV model with two decay parameters. The corresponding model 
	is given by
	\begin{equation}\label{eq:PDprofile}
	\bar{a}_n^2=\bar{a}_0^2\,\exp\left({-t_n}/{\gamma}\right),
	\end{equation}
	where $\bar{a}_0^2$ and $\bar{a}_n^2$  are the (statistical) average power of the first and $n$th multipath component, 
	respectively, and $\gamma$ is the power decay time constant for arriving rays, assumed as a random variable. 
	We estimate a $\gamma_k$ for each measurement in every scenario using a least-squares curve fitting on
	$\log(a_n^2)/\log(a_0^2)$, as shown by the examples in Fig.~\ref{fig:Gamma}.

\begin{figure}
	\centering
	\begin{subfigure}{0.4\textwidth}
	\centering
	\psfrag{Y}{\small PDF}
	\psfrag{X}{$\gamma$ (ns)}
	\psfrag{measured data}{\small measured data}
	\psfrag{Gaussian fitting}{\small Gaussian fitting}
	\psfrag{Weibull fitting}{\small Weibull fitting}
	\psfrag{Gamma fitting}{\small Gamma fitting}
	\includegraphics[width=3.2 in,height=1.5in]{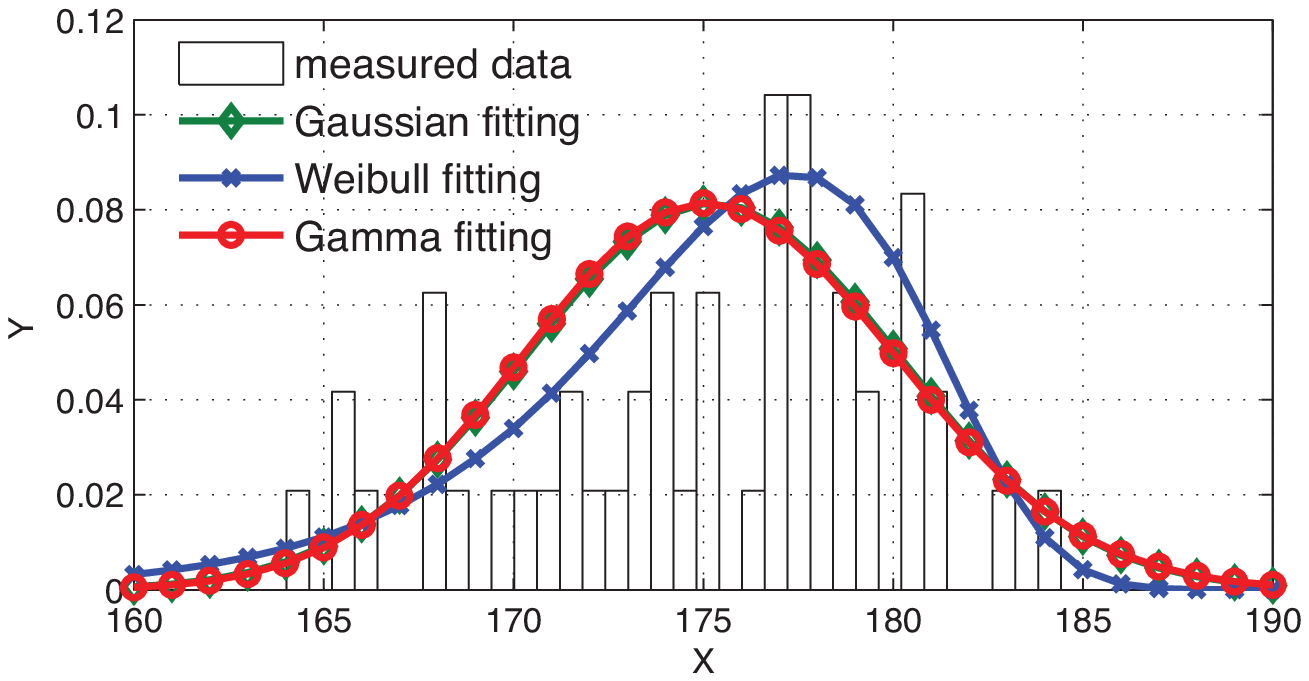}
	\caption{\it Scenario 1\\ Gaussian$(\mu,\sigma)=(175.2,4.901)$\\ Gamma$(\alpha,\beta)=(1281,0.137)$\\ Weibull
	$(\zeta,k)=(177.5,42.28)$}
	 \label{fig:GammaDis1}
	\end{subfigure}
	\begin{subfigure}{0.4\textwidth}
	\centering
	\psfrag{Y}{\small PDF}
	\psfrag{X}{$\gamma$ (ns)}
	\psfrag{measured data}{\small measured data}
	\psfrag{Gaussian fitting}{\small Gaussian fitting}
	\psfrag{Weibull fitting}{\small Weibull fitting}
	\psfrag{Gamma fitting}{\small Gamma fitting}
	\includegraphics[width=3.2 in,height=1.5in]{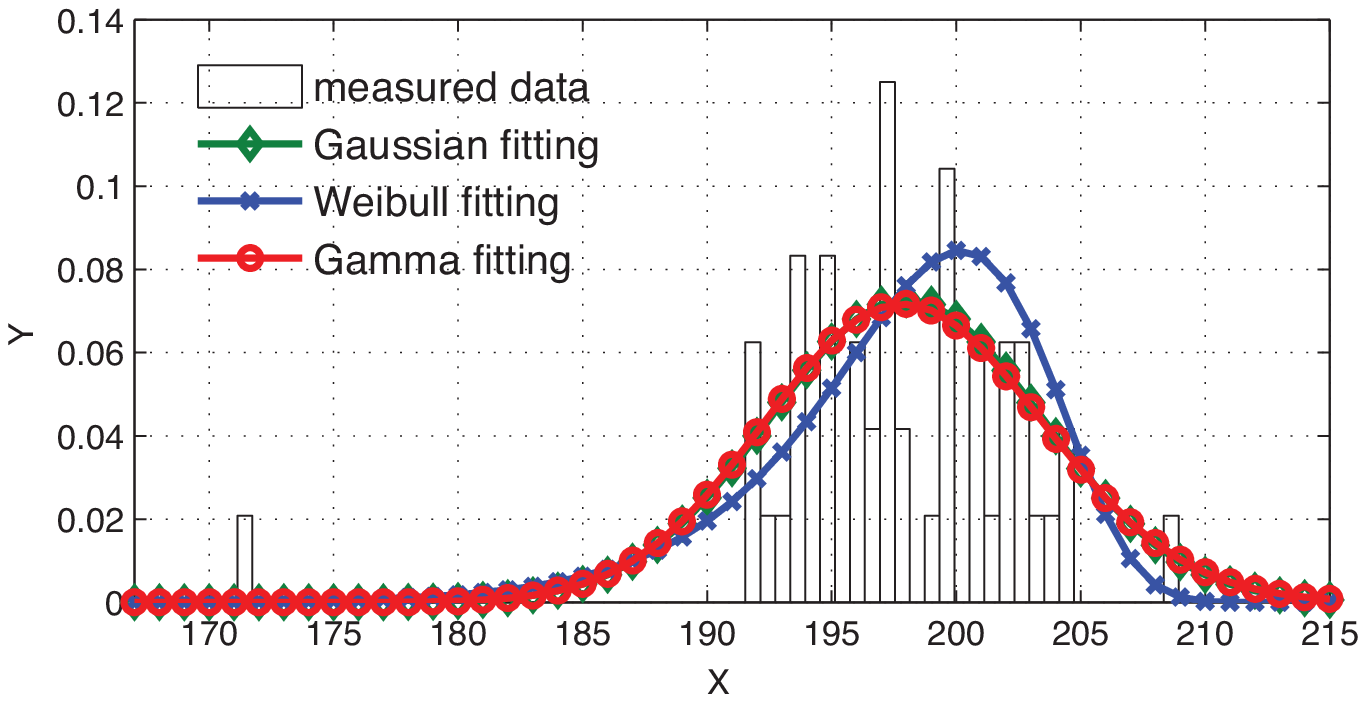}
	\caption{\it Scenario 2\\ Gaussian$(\mu,\sigma)=(197.9,5.481)$\\Gamma$(\alpha,\beta)=(1265,0.156)$\\ Weibull$
	(\zeta,k)=(200.3,46.05)$}
	 \label{fig:GammaDis2}
	\end{subfigure}
	\begin{subfigure}{0.4\textwidth}
	\centering
	\psfrag{PDF}{\small PDF} 
	\psfrag{X}{$\gamma$ (ns)}
	\psfrag{measured data}{\small measured data}
	\psfrag{Gaussian fitting}{\small Gaussian fitting}
	\psfrag{Weibull fitting}{\small Weibull fitting}
	\psfrag{Gamma fitting}{\small Gamma fitting}
	\includegraphics[width=3.2 in,height=1.5in]{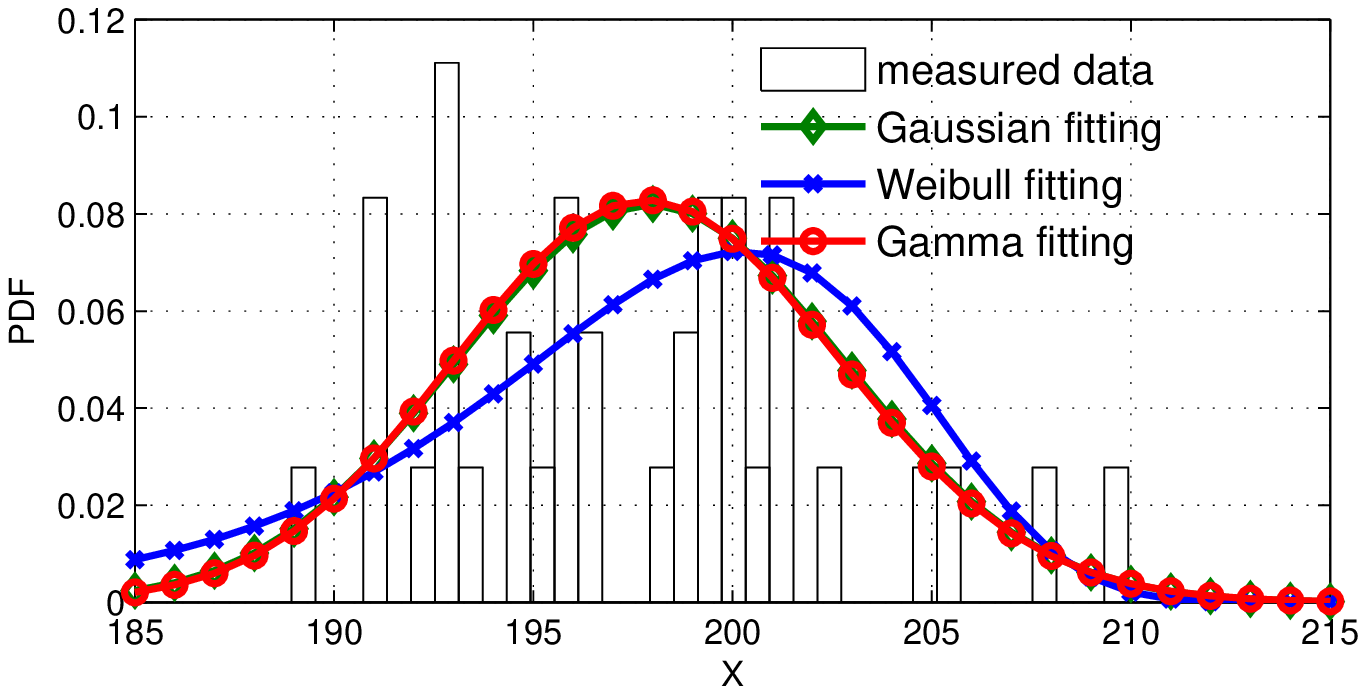}
	\caption{\it Scenario 3\\ Gaussian$(\mu,\sigma)=(197.9,4.865)$\\Gamma$(\alpha,\beta)=(1689,0.117)$\\ Weibull
	$(\zeta,k)=(200.4,39.37)$}
	 \label{fig:GammaDis3}
	 \end{subfigure}
	\caption[$\gamma$ Distribution]{PDF fittings for time decay constant $\gamma$.}
	\label{fig:GammaDist}
	\end{figure}

	Based on these estimates for the $\gamma_k$s, the PDF for $\gamma$ is
	plotted and fitted to Gaussian, Gamma, and Weibull distributions
	for each considered scenarios, as shown in
	Fig.~\ref{fig:GammaDist}.  These distributions are commonly 
	used to statistically model $\gamma$ \cite{Zoubir1,Zoubir2}.
       We use the Gaussian mean as $\gamma$ for the rest of paper.
	The Gamma distribution is given by
	\begin{equation}
	f(x|\alpha,\beta)=\frac{x^{\alpha-1}}{\beta^{\alpha}G(\alpha)}\,\exp(-\frac{x}{\beta}),
	\end{equation}
	where $G(\alpha)$ is a Gamma function, and the parameters $\alpha$ and $\beta$ are computed for all 
	scenarios from the empirical data.
	The Weibull distribution is expressed as
	\begin{equation}
	f(x|\zeta,k)= \Big\{ 
	  \begin{array}{l l}
	    \frac{k}{{\zeta}^k}\,x^{k-1}\,\exp\big(-(\frac{x}{\lambda})^k\big) & \quad \textrm{if $x\geqslant 0$ }\\
	    0 & \quad \textrm{if $x<0$ }\\
	  \end{array} 
	\end{equation}
	where the scale and shape parameters are $\zeta$ and $k$ respectively. 
	\subsection{Multipath arrival times}
	Without considering clusters, the multipath arrival times $t_n$
	would typically for indoor channels \cite{Zoubir1} be modeled as a single Poisson process.
	Accordingly, the inter-arrival times $t_n-t_{n-1}$ are modeled by an exponential PDF as 
	\begin{equation}
	p(t_n|t_{n-1})=\lambda \exp  \big( -\lambda (t_n - t_{n-1}) \big)
	\end{equation}
	where $\lambda$ is the mean arrival rate of the multipath components. It is motivated in \cite{Zoubir2} that when 
	the measured arrival times deviate too much from the single Poisson model, a mixture of two Poisson 
	processes is more suitable for modeling their arrival times. The mixture of two Poisson processes can be 
	expressed as 
	 \begin{eqnarray}
	p(t_n|t_{n-1})& = & b \, \lambda_1 \exp  \big( -\lambda_1 (t_n - t_{n-1}) \big) \nonumber \\
	&+&  (1-b) \, \lambda_2 \exp  \big( -\lambda_2 (t_n - t_{n-1}) \big)
	 \end{eqnarray}
	where $\lambda_1$ and $\lambda_2$ are the arrival rates and parameter $0 \leq b \leq 1$ is the mixing 
	probability.
	\begin{figure}
	\begin{subfigure}{0.4\textwidth}
	\centering
	\psfrag{X}{\small inter-arrival times (ns) }\psfrag{Y}{\small ln (1-CDF)}
	\psfrag{measured data}{\small measured data}
	\psfrag{single poisson process}{\small single poisson process}
	\psfrag{mixed poisson process}{\small mixed poisson process}
	\includegraphics[width=3.2 in,height=1.5in]{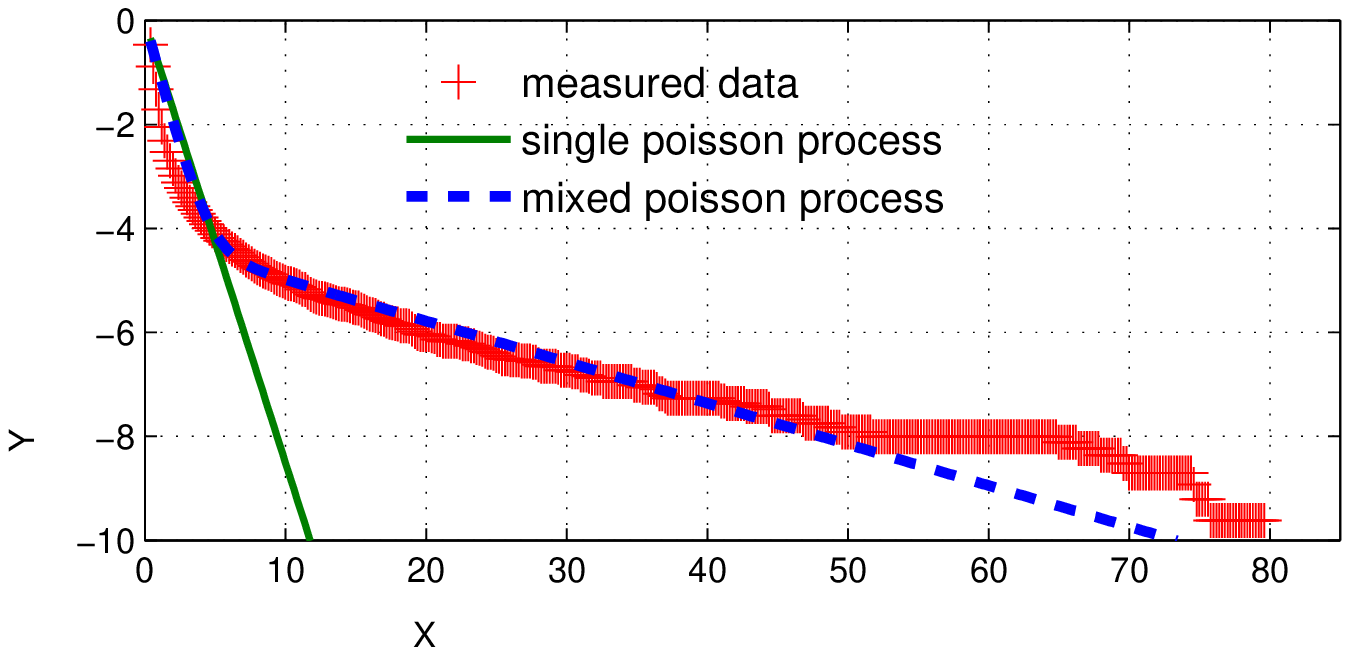}
	\caption{\it Scenario 1:\\ $\lambda=0.985, ({\lambda}_1,{\lambda}_2,b) = (0.083, 1.180, 0.015) . $}
	 \label{fig:Poisson1}
	\end{subfigure}
	\begin{subfigure}{0.4\textwidth}
	\centering
	\psfrag{X}{\small inter-arrival times (ns) }\psfrag{Y}{\small ln (1-CDF)}
	\psfrag{A}{\small measured data}
	\psfrag{B}{\small single poisson process}
	\psfrag{C}{\small mixed poisson process}
	\psfrag{X}{\small inter-arrival times (ns) }\psfrag{Y}{\small ln (1-CDF)}
	\includegraphics[width=3.2 in,height=1.5in]{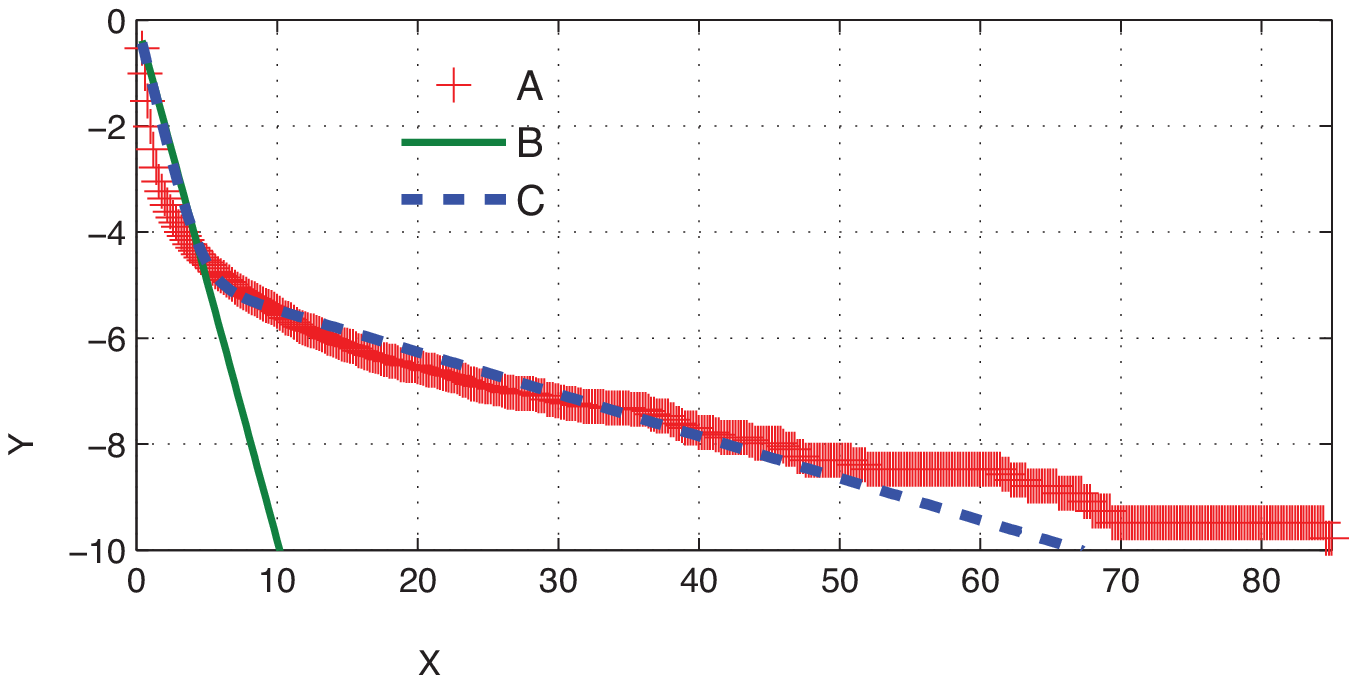}
	\caption{\it Scenario 2:\\ $\lambda=1.037, ({\lambda}_1,{\lambda}_2,b) = (0.059, 1.219, 0.008) .$ }
	 \label{fig:Poisson2}
	\end{subfigure}
	\begin{subfigure}{0.4\textwidth}
	\centering
	\psfrag{X}{\small inter-arrival times (ns) }\psfrag{Y}{\small ln (1-CDF)}
	\psfrag{measured data}{\small measured data}
	\psfrag{single poisson process}{\small single poisson process}
	\psfrag{mixed poisson process}{\small mixed poisson process}
	\psfrag{X}{\small inter-arrival times (ns) }\psfrag{Y}{\small ln (1-CDF)}
	\includegraphics[width=3.2 in,height=1.5in]{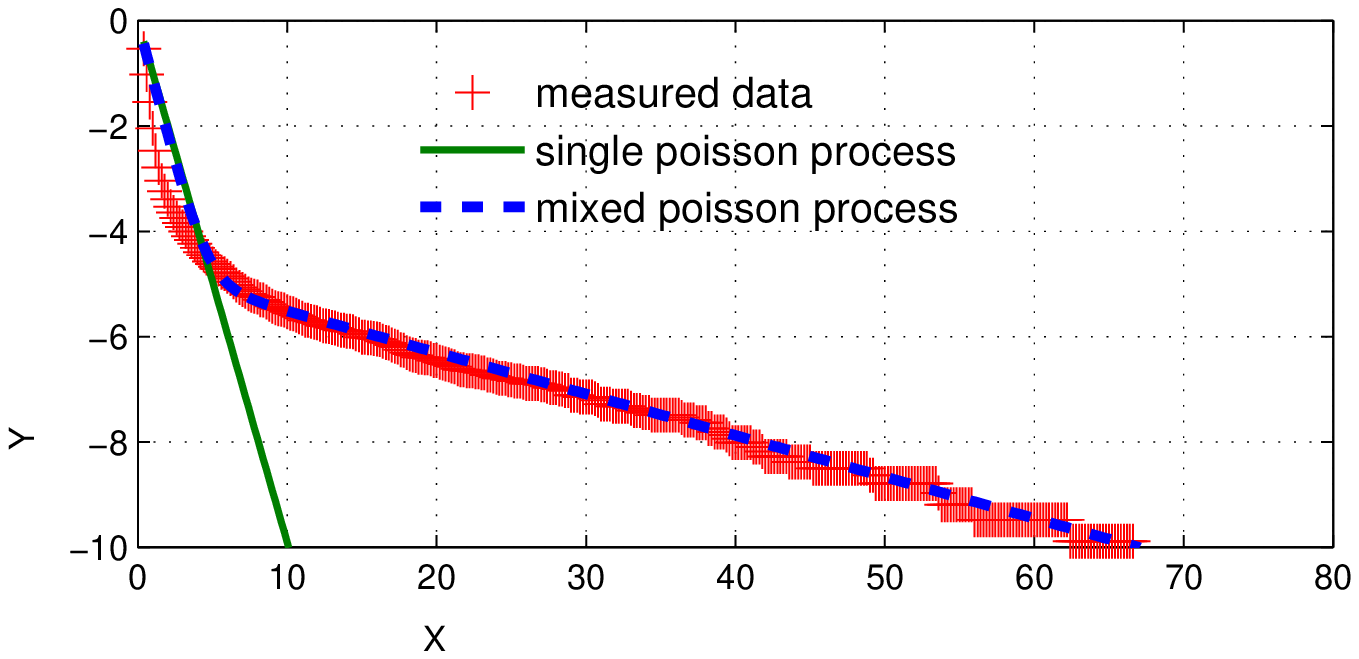}
	\caption{\it Scenario 3:\\ $\lambda=1.094, ({\lambda}_1,{\lambda}_2,b) = (0.084, 1.235, 0.009) .$}
	 \label{fig:Poisson3}
	 \end{subfigure}
	  \caption[Inter Arrival Times]{Logarithm of the complementary CDF of the inter-arrival times.}
	  \label{fig:Poisson}
	\end{figure}

	Fig.~\ref{fig:Poisson} shows
	the corresponding estimated parameters. As seen, the mixed Poisson 
	process provides a much closer fit to the measured data than the conventional single Poisson process.
	Apparently, if the RDS or channel length is large, the arriving
	paths appear over a wide range of time differences which makes
	it difficult to be represented by only one Poisson parameter.

	%
	\section{Comparison to 802.15 channel models}\label{sec:VI}

	The estimated parameters for the channel model are summarized 
	in Table~\ref{Table2}. 
	In this table, the listed parameters are:
	\begin{description}
	\item[ $P_L(d_0)$]\hspace{0.2cm}: path-loss at reference distance $d_0$~(m)
	\item[ $\alpha$]: path-loss exponent 
	\item[ $\sigma_{P_L}$]:  path-loss log-normal standard deviation 
	\item[ $\bar{L}$]: mean RMS delay spread (RDS)
	\item[ $\Lambda$]: cluster arrival rate 
	\item[ $\lambda$]: ray arrival rate (single Poisson fit)
	\item[ $\Gamma$]:  power decay constant for clusters
	\item[ $\gamma$]: power decay constant for rays
	\item[ $\sigma_{\Gamma}$]: cluster power decay log-normal standard deviation
	\item[ $\sigma_{\gamma}$]: ray power decay log-normal standard deviation
	\end{description}

	Also, the corresponding parameters of the standard IEEE 802.15 channel model are shown.
	The numbers are taken from \cite{TG3report}, which 
	provides models for wideband (9~GHz bandwidth) channels at 60~GHz carrier frequency. 
	The reported parameters are selected from the CM1, CM4, and CM9 channel models suggested in 
	this document and obtained from the measurements in residential LOS environment, 
	office area in NLOS scenario, and within a kiosk with LOS, respectively.

	\begin{table*}
	\centering
	    \begin{tabular}{cc|ccc|ccc|}
	     \cline{3-8}
	                      &       & \multicolumn{3}{c}{Metal cabinet} & \multicolumn{3}{c|}{IEEE 802.15}\\ \cline{3-8}
			      &       & Sc. 1&  Sc.2 & Sc. 3 & CM1 & CM4 & CM9\\  \cline{1-2}
	    \multicolumn{1}{|c} {Parameter} &  \multicolumn{1}{c|} {  Unit}    & LOS  &  LOS  & NLOS  & LOS & NLOS & LOS\\
	    \hline
	   \multicolumn{1}{|c} {$P_L(d_0)$}& dB& 54.711 & 53.439 & 54.116 & 75.1 & 56.1 & NA\\ 
	   \multicolumn{1}{|c} {$\alpha$}         &      & 0.02 & 0.004 & 0.002 & 1.53 & 3.74 &NA \\
	    \multicolumn{1}{|c}{$\sigma_{PL}$}    & dB   & 0.39 & 0.17 & 0.16 & 1.5 & 8.6 & NA\\
	    \multicolumn{1}{|c}{$\bar{L}$}        & ns   & 113.4 &  158.3 &  159.1 & NA & NA & NA\\
	    \hline
	    \multicolumn{1}{|c}{$\Lambda$}        & 1/ns & - & - & - & 0.144 & 0.07 & 0.044\\
	    \multicolumn{1}{|c}{$\lambda$}        & 1/ns &0.985 &  1.037 & 1.094 & 1.17 & 1.88 & 1.01\\
	    \multicolumn{1}{|c}{$\Gamma$}         & ns   & - & - & - & 21.5 & 19.44 & 64.2\\
	    \multicolumn{1}{|c}{$\gamma$}         & ns   & 175.23 & 197.99 & 197.93 & 4.35 & 0.42 & 61.1\\
	    \multicolumn{1}{|c}{$\sigma_{\Gamma}$}& ns   & - &  - & - & 3.71 & 1.82 & 2.66\\
	    \multicolumn{1}{|c}{$\sigma_{\gamma}$}& ns   & 4.90 &  5.48 &  4.86 & 7.31 & 1.88 & 4.39\\
	    \hline
	     \end{tabular}
	    \caption{Comparison of various channel parameters of the measured channels, compared to IEEE 802.15 channel models.
	    Abbreviation ``NA" stands for not available and ``-" means not applicable here.  } \label{Table2}
	\end{table*}

	It can be seen from  Table~\ref{Table2} that the measured channel in a metal enclosure differs significantly 
	from the typical wireless channels that are used in the literature to design a wireless link for short range 
	communications.  The main distinctions are:
	\begin{enumerate}
	\item Very small path-loss exponent in both LOS and NLOS cases,
	\item Significantly longer channels or equivalently very large RDS,
	\item Arriving rays do not form clusters,
	\item Arrival rate is modeled here as a mixed Poisson process.
	\end{enumerate}
	
	Obviously, the very small path-loss exponent and resulting very large RDS are caused by the non-damping
	effect of the metal walls.  The resulting long channels in metal enclosure environment make wireless system design
	and the involved signal processing both an interesting as well
	as a challenging task.

	%
	\section{System design example}\label{sec:VII}
	Channel models are generally used for system design. For the presented model of a metal enclosure at 60~GHz, 
	a distinguishing feature is the very large delay spread. Correspondingly,
	the rich multipath makes the wireless channel very frequency selective, as is clear in Fig.~\ref{fig:CFR}.
	For such a channel, OFDM is an appropriate modulation scheme,
	and is indeed considered for most of the existing wideband
	wireless standards including WiMAX, LTE, WiFi, and also for the
	upcoming new standard for 60 GHz WPAN, i.e., IEEE 802.15.3c.  In
	the usual designs, the frequency band is divided into several
	subcarriers such that each subcarrier experiences a flat-fading
	channel.

	In this section, we compare the performance of an OFDM system using the measured channel to that using a
	simulated Rayleigh fading channel. This provides directions and questions for further work, such as
	\begin{itemize}
	\item What are suitable modulation schemes and equalization techniques for such dispersive (rich 
	scattering) and extremely long channels? 
	\item Can we benefit from the diversity gain offered by these types of channels with an acceptable computational 
	complexity or should the channel be shortened using e.g.\ absorbers?
	\item What is the channel capacity for such highly reflective environments?
	\end{itemize}
	For the numerous existing wireless channel models, these questions are well studied.
	The initial approach would be to relate the proposed channel model to the available models and 
	modify the system design including the modulation,  coding and equalization in order to cope with the new 
	circumstances.
	\subsection {Design parameters}
	%
	
	%
	\begin{figure}
	\centering
	\psfrag{P}{\small pilot}
	\psfrag{D}{\small data}
	\psfrag{E}{\small frequency}
	\psfrag{L}{\small guard band}
	\psfrag{T}{\small time}
	\psfrag{G}{\small guard band}
	\psfrag{B}{\small $N_u$}
	\psfrag{A}{\small $N_{cp}$}
	\psfrag{C}{\small $N$}
	\psfrag{F}{\small IDFT}
	
	\includegraphics[width=3.3 in,height=1.5 in]{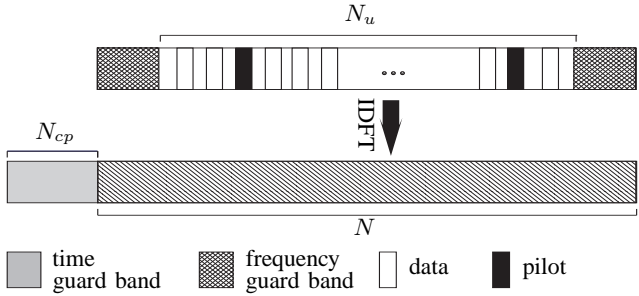}
	\caption[CDF of RDSs]{ data sequence with frequency guard bands (null subcarriers 
	in frequency domain) and pilot subcarriers is converted to the time domain OFDM block by
	taking IDFT and appending the time domain guard (cyclic prefix) to the block.}
	 \label{fig:OFDMblock}
	\end{figure}

	The design example is based on a single antenna, standard OFDM modulation without coding (see Fig.\ \ref{fig:OFDMblock}).
	An OFDM block with bandwidth $B_w$ is split into $N$ subcarriers, consisting of guard bands and $N_u$ 'user' subcarriers 
	(data and pilots). In time domain, the corresponding $N$ samples are augmented with a cyclic prefix of $N_{cp}$ samples.
	The symbol duration is $T_s = 1/B_w$.
	Straightforward equalization requires that the time duration of the cyclic prefix is larger than the length of the wireless channel:
	\begin{equation}\label{eq:OFDMdesign1}
	    N_{cp} T_s > t_{max}  \quad 
	\end{equation} 
	This is used to isolate the ISI within each block of the OFDM symbol so that the ISI can be eliminated separately in each block 
	by frequency domain equalization. 

	We assume $M$-ary modulation, with $M = 2^l$, so that a symbol consists of $l$ bits. 
	The bandwidth efficiency is
	\[
	\kappa = \frac{l\,N_u}{N_{cp}+N}
	\]
	and the resulting data rate is $\kappa/T_s$ bits per second.  The data rate can be increased by increasing $l$ but this will reduce
	the SNR and lead to a higher bit-error rate (BER). We can also increase $N_u$ (hence also $N$) until the bandwidth efficiency
	saturates to $l$. Finally, the symbol duration $T_s$ can be shortened by increasing the available bandwidth $B_w$.

	Latency is often also a consideration, and this leads to a limitation of the size of a data packet. One OFDM block will have a
	duration of
    \[
        (N_{cp} + N) T_s
    \]
        and some systems pose a maximum to this.

	Several other limitations are in place.  
	Apart from practical limitations and computational limitations,
	the number of subcarriers ($N$) that can be allocated in one
	OFDM block is limited by the requirement that the channel is
	constant over the duration of the OFDM symbol, i.e., the
	coherence time of the channel should be larger.  The coherence
	time is defined as $\frac{1}{\Delta f_D}$ where $\Delta f_D$ is
	the range of possible Doppler frequencies of the channel, and
	the requirement becomes \cite{Goldsmith}
	\begin{equation}\label{eq:OFDMdesign2}
	(N+N_{cp})\,\Delta f_D\,T_s \ll 1\,,
	\end{equation} 
	Another requirement is that each subcarrier experiences flat fading. 
	In frequency domain, the distance between fades is related to
	$1/t_{max}$. This leads to \cite{Goldsmith}
	\begin{equation}\label{eq:OFDMdesign3}
	    N \gg B_w t_{max}
	\end{equation} 

	As an example, let us design a system at $f_c = 60$ GHz with an available bandwidth $B_w = 5$ GHz.
	For the sake of simplicity, we consider BPSK modulation, which leads to $l=1$ and $T_s=\frac{1}{B_w}=$0.2~ns.

	The proposed system is part of a mechatronic system in a closed metal environment in which a moving platform with sensors and
	actuators has to communicate to a controller in the ``fixed'' world.
	Since movements that occur outside the enclosure do not affect the channel, we expect a slowly 
	time-varying channel with a sufficiently long coherence time.
	The Doppler shift is defined as $\Delta f_D=\frac{\nu f_c}{c}$, where $\nu$ is the relative speed between 
	transmitter and receiver, $c$ is the speed of light, and $f_c$ is the carrier frequency. If we assume a maximum
	relative speed of $10~\mathrm{ms}^{-1}$, then the Doppler frequency range is $\Delta f_D = 2$~kHz, and
	the coherence time of the channel is $\frac{1}{\Delta f_D} = 0.5$~ms.
	
	As discussed earlier, the spectral efficiency increases with the number of subcarriers. Considering
	\eqref{eq:OFDMdesign2} for $\Delta f_D T_s=0.4 \cdot 10^{-6}$, the upper bound 
	on the length of an OFDM block is given by $N_{cp}+N \ll 2.5\cdot 10^6$.
	To satisfy this, we consider as constraint on the size of a transmission block
	\[
	N_{cp}+N < 2.5 \cdot 10^5\;\; \mbox{symbols}. 
	\]
	The cyclic prefix $N_{cp}$ should satisfy (\ref{eq:OFDMdesign1}). For a channel of length $t_{max} = 1 \mu$s, this leads to 
	\[
	N_{cp} \ge 5000 \,.
	\]
	The exact number depends on the scenario.  Finally, condition (\ref{eq:OFDMdesign3}) leads to
	\[
	N \gg 5000 \,.
	\]

	We consider two versions of the system: (A) minimal latency; (B)
	maximal data rate.  For minimal latency, we take $N_{cp}=5000$, 
	and a block size of $N=2^{13}=8192$ of which we
	take $N_u=6720$ data/pilot symbols, and $2\times
	N_{\text{guard}}= 1472$ null subcarriers for frequency guards at
	both ends of a block. This leads to a spectral efficiency of $\kappa \approx
	0.5$.  The duration of one data packet becomes
	2.64~$\mu$s and the data rate is 2.547~Gbps.

	For maximal data rate, we choose $N = 2^{17} \approx 1.3 \cdot
	10^5$, of which we take $N_u = 107520$ data/pilot
	symbols, and $2\times N_{\text{guard}}= 23552$ null subcarriers.
	The duration of one data packet is about 27.2~$\mu$s and the data rate
	is 3.95~Gbps. Thus, latency increases almost 10 times by taking $N = 2^{17}$
	compared to  $N=2^{13}$ where as the data rate increases by less than 50\%.

	For the measured channels, the channel length $t_{max}$ varies depending on the scenario, which alters 
	the cyclic prefix and consequently the block length.
	Table \ref{Table2} summarizes the parameters that are used in the simulation.
	As usual, higher data rates could be obtained by considering higher-level modulation as well as multiple antennas.
	\begin{table*}
	\centering
	    \begin{tabular}{ ll | ll | ll  |}
	     \cline{3-6}
	     &&\multicolumn{2}{c}{ A: $N = 2^{13}$}  &  \multicolumn{2}{ c |}{ B: $N = 2^{17}$}\\ \cline{2-6}
	     &  \multicolumn{1}{|c|}{$N_{\text{cp}}$}  & bit-rate (Gbps) & Latency ($\mu$s  ) & bit-rate (Gbps) & Latency ($\mu$s  )  \\ \hline
	    \multicolumn{1}{|c|}{Simulated 1}  & 11 & 4.096&  1.640         &4.101     &26.216    \\
	    \multicolumn{1}{|c|}{Simulated 2}  & 4001 & 2.756& 2.438          &3.980     &27.015    \\
	    \multicolumn{1}{|c|}{Scenario 1} & 3903 & 2.778 &  2.418         &3.983     & 26.994   \\
	    \multicolumn{1}{|c|}{Scenario 2} & 5812 & 2.399 &  2.801         &3.927     &27.377    \\
	    \multicolumn{1}{|c|}{Scenario 3} &5617  & 2.433 & 2.762          &3.933     &27.338    \\  \hline
	    \end{tabular}
	    \caption{Data rates and latency for (A) a low-latency system, and (B) a high-rate system.} \label{Table2}
	\end{table*}

	\subsection {Simulation setup} 

	For the designed BPSK-OFDM systems, we will compare
	the BER performance for both simulated Rayleigh fading channels and the measured channels for the various scenarios.
	The Rayleigh fading channel is based on a tapped delay line setup where each path is 
	assumed to be a Rayleigh fading process without considering any specific power delay profile. 

	The considered performance measure is BER as function of $E_b/N_0$, where $E_b$ is the transmit energy per bit,
	and $\frac{N_0}{2}$ is the two-sided noise power spectral density (PSD). 
	In the simulation, we first convert $E_b$ to the energy per symbol $E_s$, taking into account the number of bits per symbol $l$
	and the overhead by the cyclic prefix $N_{cp}$, resulting in
	\begin{equation}\label{eq:EbN0}
	\frac{E_s}{N_0} = \frac{E_b}{N_0} \cdot \frac{l N_u}{N_{cp}+N} \,.
	\end{equation}
	The measured and simulated channels are normalized to unit power and
	convolved with the transmit sequence.  In that case, the transmitted
	$E_s/N_0$ is equal to the received signal to noise ratio (SNR), and
	we add white Gaussian noise of suitable power to obtain the
	specified SNR.

	For such dispersive channels, several subcarriers experience
	fading so that the BER is usually not very good.  Various
	well-known techniques could be introduced to combat the fading
	subchannels, e.g., channel coding, interleaving, rake receiver
	design, as well as single input multiple output (SIMO) systems
	with diversity combining schemes \cite{Goldsmith}.  However, for
	improved interpretation of the results, we will not consider
	these in the simulation.

	In the simulation, we assume that the receiver has perfect knowledge of the channel.

	\begin{figure}
	\centering
	\psfrag{Y}{\small BER}
	\psfrag{X}{\small $E_b/N_0$}
	\psfrag{Theory}{\small Rayleigh (theory)}
	\psfrag{measred 11}{\small measured channel scenario1 ($N=2^{13}$)}
	\psfrag{measured12}{\small measured channel scenario1 ($N=2^{17}$)}
	\psfrag{measured 31}{\small measured channel scenario3 ($N=2^{13}$)}
	\psfrag{measured21}{\small measured channel scenario2 ($N=2^{13}$)}
	\psfrag{Simulation1}{\small simulated short Rayleigh channel  ($N=2^{17}$) }
	\psfrag{Simulation2}{\small simulated long Rayleigh channel  ($N=2^{17}$)}
	\includegraphics[height=3.2in,width=3.2 in]{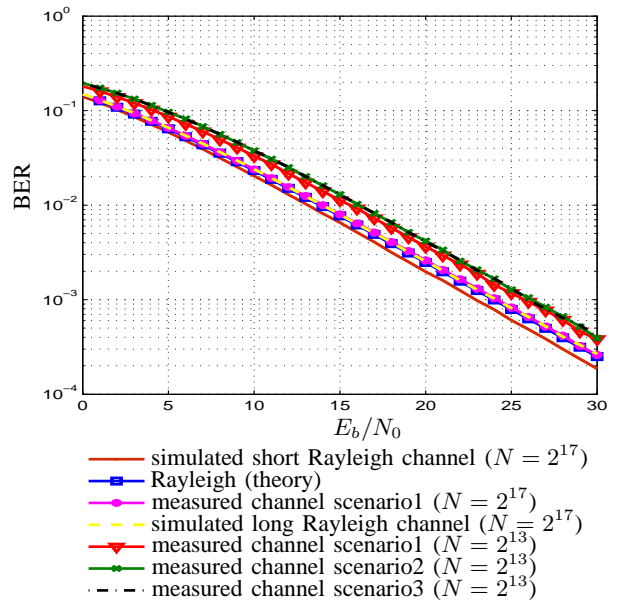}
		  \caption[Measurement Setting 1]{BER performance of an uncoded OFDM  system over
	          the measured channels for different scenarios with block size $N=8192$
	          and $N=2^{17}$  .
	          The BER plot of OFDM block of size  $N=2^{17}$ is shown for {\it scenario1}, 
	          ({\it scenario2} and {\it 3} have the same BER curves as {\it scenario1}), and
	          also for two simulated Rayleigh fading channels with $L=10$ and $L=4000$ taps.
	          A theoretical BER curve for a narrowband signal (No ISI) is plotted as a reference.}
	 \label{fig:BERplot}
	\end{figure}

	\subsection {Simulation results}
	
	Fig.~\ref{fig:BERplot} shows the BER as a function of $E_b/N_0$ for the various channels.
	As expected, the performance is generally limited by the fading channel as the symbols in 
	channel nulls cannot be recovered by frequency domain equalization. 
	For the larger block size (design (B)), the BER performance is slightly (almost 3~dB) better than the shorter 
	OFDM block, as expected, due to the spectral efficiency and less spread of transmit power over the cyclic symbols.

	%
	\section{Summary and Concluding Remarks}\label{sec:VIII}
	In this paper, a comprehensive channel model is provided for 60~GHz transmission inside a metal 
	enclosure based on channel measurement results for different scenarios including the LOS and NLOS 
	situations. Significantly long channels together with very rich multipath reflected from the metal walls are the 
	most distinguishing features of such channels.
	Both large-scale and small-scale channel characteristic parameters are compared with other 60~GHz 
	modeling results for short range wireless communication which indicates a noticeable difference between 
	transmission in a metal surrounding and conventional indoor environments.
	
	An uncoded OFDM system design example is elaborated, and the BER
	performance is studied using the measured channels, which
	identifies the fading effect on the system functionality.  We
	considered OFDM because time domain equalization is
	computationally expensive for such long channels.  
	Moreover, to combat the poor BER due to
	frequency-domain fades, some form of coding should be used,
	which further reduces the bandwidth efficiency.  To limit the
	loss of data rate due to the long cyclic prefix, channel
	shortening or per-tone equalization techniques
	\cite{Geert1,Barhumi} could be used.

	The system design example shows that high data rates and reasonable latencies are feasible, thus enabling interesting
	applications in demanding mechatronic systems.  The computational complexity will have to be verified. 
	Further data rate improvements are possible by exploiting the rich scattering environment in combination with large MIMO antenna
	arrays (that will be compact at 60 GHz), which will also boost the SNR due to improved spatial diversity.
	The latency of the system is determined by the long channel impulse response and the long cyclic prefix.
	If latency is an issue, physical remedies can be suggested
	including an absorbing coating inside the metal enclosure
	\cite{absorber60}, if restrictions on installing such bulky
	materials inside the mechatronic system permits this.

	%
	\section*{Acknowledgment}
	We would like to thank Dr.~Marco Spirito and the staff of the Electronics Research Laboratory at TU Delft for 
	providing the channel measurement setup.
	%
	\bibliographystyle{IEEEtran}
	\bibliography{IEEEabrv,arXivNovSeyran}
	\end{document}